Galactic Atmospheres

# A Practical Guide to Hosting a Virtual Conference

Cameron Hummels[1] Benjamin Oppenheimer[2] G. Mark Voit[3] Jessica Werk[4]

[1]California Institute of Technology, [2]University of Colorado, Boulder, [3]Michigan State University, [4]University of Washington

Virtual meetings have long been the outcast of scientific interaction. For many of us, the COVID-19 pandemic has only strengthened that sentiment as countless *Zoom* meetings have left us bored and exhausted. But remote conferences do not have to be negative experiences. If well designed, they have some distinct advantages over conventional in-person meetings, including universal access, longevity of content, as well as minimal costs and carbon footprint [1]. This article details our experiences as organizers of a successful fully virtual scientific conference, the KITP program "Fundamentals of Gaseous Halos" hosted over 8 weeks in winter 2021. Herein, we provide detailed recommendations on planning and optimization of remote meetings, with application to traditional in-person events as well.  We hope these suggestions will assist organizers of future virtual conferences and workshops.

Our experiences are derived from [Fundamentals of Gaseous Halos](), an 8-week scientific workshop hosted by the [Kavli Institute for Theoretical Physics]() (KITP). Also known as **Halo21**, this program brought together specialists from the numerous, disparate communities of observers and theorists whose research involves the tenuous gas surrounding galaxies known as the circumgalactic medium (CGM). **Halo21** was originally planned to be hosted in-person, but due to the constraints of the COVID-19 pandemic, we transformed it into a fully online program. We received overwhelmingly positive feedback from the community throughout the workshop, and we incorporated participant suggestions as the program unfolded to further improve the quality of the program.

Below we recommend some specific approaches for virtual meeting organizers to consider as they plan their own virtual scientific workshops and comment on the implementation and efficacy of those approaches during the **Halo21** experiment.

## 1) Choose a Thematic Structure for the Program

Our first recommendation is not unique to virtual programs.  Choose a thematic layout for the sessions of your program, so that activities and speakers addressing similar topics are grouped together.  Those topics can be ordered so that they define an overall narrative for the meeting.  Providing such a framework will benefit your audience by ensuring that the participants are focusing on one topic at a time, so that they can explore it in greater depth and introspection.  For virtual meetings in particular, a structured approach to organization allows your participants to set aside blocks of time for the topics that interest them the most, if other commitments prevent them for attending the entire meeting.  We recommend keeping these themes as simple and concise as possible.

### Comments based on *Halo21*:

We organized **Halo21** around eight key questions in CGM science, addressed at a rate of one per week. Those questions were quite general and non-technically phrased.

- Week 1: How do gaseous halos depend on halo mass?
- Week 2: Why are gaseous halos often multiphase?
- Week 3: What roles do non-thermal components play?
- Week 4: What does the Milky Way tell us?
- Week 5: How does gas flow out of galaxies?
- Week 6: How does gas flow into galaxies?
- Week 7: How do gaseous halos affect galaxy evolution?
- Week 8: What future observations will transform our understanding?

Each week featured two keynote speakers who were asked to focus on the week's key thematic question, rather than simply presenting a summary of their own latest research contributions. Those talks therefore provided background and structure that engaged all participants more broadly than a narrow look at recent topical developments. We found that this focus on big questions laid a foundation that made each week's discussions more inclusive and beneficial, especially for participants with research interests and background different from those of the keynote speakers.

## 2) Choose Meeting Times that Work for Your Audience

Whether your virtual conference lasts a week or extends over multiple months, you will need to choose appropriate times for live sessions, during which the participants synchronously interact. The times you choose will impact what parts of the world will be able to participate in real time and will favor participants for whom your meeting time coincides with a usual weekday work day (9AM-5PM local time). Furthermore, consider the daily duration of synchronous meeting time for your audience—are participants going to spend their entire work day on your program during a whole week, or are they trying to continue with their normal career obligations while fitting in a few program hours once per day or maybe every other day? Beware of overly long days of engagement that burn out your participants and organizers alike!

### Comments based on *Halo21*:

We chose to hold live sessions for two hours each day during an eight-week period, which allowed participants to keep up with their normal weekly obligations and avoided burnout. Most scientific meetings do not extend over eight weeks like a KITP program does, but this feature greatly broadened participation because of how it allowed people with crowded schedules to keep pace with the meeting. We suggest that others planning virtual meetings consider a limited daily schedule spread over a longer time period than a typical in-person meeting, as your participants aren't limited by travel and funding needs.

Our daily meeting time was 8-10AM Pacific Time, which enabled reasonable connection hours spanning the Americas, Africa, and Europe but was an obstacle for participants wanting to join from Asia, Australia, and Oceania. We considered shifting around the meeting time over the course of the meeting to accommodate different time zones, but in the end decided that would be too disruptive after people had adapted to the pattern.

Participation numbers varied over the course of the week or the program, depending on everyone's personal interest and obligations. However, video recordings of each day's activities (see #6) and asynchronous communication afforded by *Slack* (see #3) mitigated many of the problems of time zones and people skipping days here and there.

## 3) Use Both Synchronous and Asynchronous Platforms

Most conferences focus on synchronous interaction through presentations, Q&A panels, and discussions. In a virtual environment, these synchronous activities employ video-conferencing software like *Zoom*, *Google Meet*, *Skype*, and more.

One downside to a virtual meeting is that participants lack easy ways to communicate, outside the temporary and clunky interfaces of those teleconferencing applications. Fortunately, software like *Slack* can enhance communication throughout a virtual meeting (and even beyond it) in several beneficial ways:

- channel-based discussions to sequester different topics in different channels
- asynchronous communication, so if a participant missed a live session, they are still able to catch up and participate in the discussion
- threaded conversations, which can occur organically off of any comment in a discussion
- integration of links, images, and movies directly into conversations, which greatly aids in linking to relevant papers and plots in discussions
- private messaging between users, allowing for the organic creation of private conversations and collaboration groups
- semi-permanent records lasting for years in the future, or as long as you are willing to host the *Slack* workspace (requires paid *Slack* subscription)

### Comments based on *Halo21*:

*Halo21* used *Zoom* as its primary face-to-face synchronous communication tool. It generally worked well, but the **Halo21** *Slack* space may have been more integral to the success of the program. In particular, it gave voice to a much larger contingent of participants than just the people saying things during the *Zoom* meetings. It also enabled hundreds of organically arising scientific conversations to spin off of the synchronous sessions, many of them enhanced with links to published work, preprints, unpublished plots, and animations.

The **Halo21** *Slack* workspace remains active as of this writing, almost four years after the start of our program. It is perhaps the program's most lasting contribution to the community, preserving hundreds of pages of detailed, annotated discussions between the participants in our workshop. Several collaborations that formed during **Halo21** have continued to use this *Slack* workspace as their primary method of communication. The only major downside to using *Slack* is that it's not free, requiring funding to preserve these discussions.

## 4) Limit Formal Presentations and Increase Interaction

How many of us have found ourselves a few days into a conference or workshop, so burned out from hearing talk after talk that we barely absorb anything we are hearing? Formal presentations can be informative at providing context, but active discussion is oftentimes where science really gets addressed [2]. Crucially, organizers must not only provide time for discussion and interaction, but actively encourage it to continue beyond the strict confines of the workshop.

### Comments based on *Halo21*:

*Halo21* integrated ample time for discussion into every aspect of the program. A few examples include:

- **Panel Discussions following Keynote Talks.** Twice a week, invited speakers presented 60-minute keynote *Zoom* presentations focusing on the week's key question, which provided essential context and broad background for the program's participants. Following each keynote talk, we organized a ~60-minute panel discussion featuring the keynote speaker alongside several other experts in the field able to provide alternate perspectives and some counterpoint. This panel was moderated by the organizers, but it accepted input and questions from the full **Halo21** community via *Slack* and *Zoom*.
- **Discussion following Tutorials.** After each of the tutorial breakout sessions (see #5), we provided substantial periods where those present could ask questions and discuss the relevant content presented during the tutorial. Since these tutorials were primarily aimed at introducing new methods and techniques, allowing interaction was crucial in ensuring people truly learned what they needed to employ these methods in their own scientific practice.
- **Slack Conversation Groups.** We encouraged participants to initiate *Slack* channels dedicated to particular subfields of CGM science, while we also seeded the *Slack* space with channels devoted to each of the weekly thematic questions (e.g., "why are gaseous halos often multiphase"). This particular feature of the workshop may have the most lasting benefits, as the many lively and insightful discussions that ensued have been preserved for others to read beyond the conclusion of the program.
- **Featured Conversations.** Each week, we designated two or three *Slack* Conversation Groups as "Featured." These conversation initiators presented general background to all program participants during Monday's *Zoom* session and returned on Friday's *Zoom* session to present a summary of notable insights and outcomes from their *Slack* conversations throughout the week.

## 5) Provide Tutorials on Fundamental Ideas, Software, and Techniques

For researchers seeking to jump into new subfields, progress can be challenging without adequate introduction to the themes and techniques intrinsic to those disciplines [3]. But virtual workshops provide an opportunity to give the necessary background on fundamental ideas, software, and techniques necessary to getting up to speed in a new subfield through weekly optional tutorials.

### Comments based on *Halo21*:

Wednesdays featured (3-4) 20-minute tutorials designed to facilitate boundary crossing among researchers in adjacent sub-fields. These *Zoom* sessions included post-tutorial breakout rooms, so that follow-up discussions could continue without delaying the start of the next tutorial.  Tutorials included (a) introductions on how to use certain types of software packages relevant to the subfield (e.g., synthetic spectral generation software); (b) step-by-step descriptions on using scientific techniques (e.g., observational measures of halo gas turbulence); (c) introduce a specific sub-discipline (e.g., physics of cooling flows).  While tutorials were enormously popular for audience members trying to learn the background on related fields, care was needed to ensure that tutorial presenters gave a broad perspective rather than a job talk highlighting only the work they themselves had done in a sub-field.

## 6) Utilizing Video Sharing Platforms for Science Content

Video-sharing platforms have taken over the internet, and their success is linked to the brevity of their content.  We as scientists can benefit from this trend too, as these videos can provide a brief and accessible means to describe paper results, present a new scientific concept, or introduce junior scientists to the community.  And with video sharing platforms like *YouTube* and *Vimeo*, this valuable content can be stored and shared indefinitely.

### Comments based on *Halo21*:

We experimented with this video-sharing format by creating a **Halo21** [YouTube channel](#) featuring all of the content from the workshop.

- **Conference Content Preserved Indefinitely.** We uploaded all of the recorded keynote lectures, tutorials, and panels from the program continuously throughout the program.  This greatly aided in people staying up to date on the program even if they missed a day or a week and needed to catch up.

- **New Results Videos.** We invited participants to share short video presentations (< 5 minutes) about recent work.   Users were invited to produce their own videos (using *Zoom* or other video capture programs) describing a paper or a result that they had written up in the previous year.  We provided a detailed tutorial on how to even create a video of this kind in *Zoom*, so users could narrate a walk through a paper alongside plots and images. These new results videos were particularly popular among the more junior participants, who otherwise might not have gotten a chance to present their findings to the larger group.  Each week we would receive 5-10 new results videos from participants, highlighting their new scientific results.  Someone could sit down and get "up to speed" in the field over the course of 30-minutes of watching videos each week.

## 7) Enable Social Engagement between Participants

One of the major downsides to virtual participation is the lack of opportunities to meet new people and to engage with colleagues socially. But virtual conferences can create specific activities for cultivating new contacts and social interaction, which can help mitigate this problem.

### Comments based on *Halo21*:

We experimented a great deal in providing informal activities for social engagement, including:

- **Pub Trivia Nights.** In-person conferences allow for participants to continue socializing after the scientific discussions of the day over a drink, so why not virtual meetings as well? We scheduled three pub trivia nights, hosted over *Zoom*, where we hired a professional "quizmaster" to test our knowledge on a variety of topics. Participants were encouraged to bring spouses or family members to join in the fun. Drinking was allowed by obviously not required. There were modest prizes (e.g., Amazon gift cards) for the winning team each week. This worked better than anticipated with 20+ enthusiastic participants each session.
- **Astronomy on Tap**. Astronomy on Tap is a is a global phenomenon where professional astronomers give informal science talks in local bars with accompanying pub trivia and interactions with the public. We decided to attempt this virtually and give public outreach talks related to the content of the *Halo21* program, also allowing for participants to join in addition to members of the public. Each event consisted of two 20-minute talks on a topic related to the *Halo21* program given at a public level. This was delivered as part of the Astronomy on Tap, Los Angeles chapter, which is run by one of the conference organizers and already had a large public following. Afterwards, we answered questions from the audience, and hosted astronomically-themed pub trivia. Each of these two sessions had ~100 attendees and have since received ~2k views on *YouTube*.

## 8) Furnish Opportunities for Junior Scientists and Networking

In-person conferences provide opportunities for junior scientists to network with the established members of a field whether by presenting on their research or through social interactions. Virtual meetings can be a real challenge for scientists new to a community because these interactions are harder to achieve. Care must be taken to ensure that junior scientists can achieve visibility and network with an existing community through both formal and informal means.

### Comments based on *Halo21*:

We made a significant effort to provide networking opportunities for both junior and senior scientists through *Halo21*:

- **Speed Collaboration.** This feature was explicitly designed to facilitate the random and sometimes serendipitous conversations that naturally happen at an in-person meeting but would not otherwise happen in a virtual one. In the second hour of each week's Monday *Zoom* session, we used breakout rooms to pair participants randomly for a 5-minute period, during which they were asked to briefly introduce their interests, seek out the common ground of their research and, if possible, identify a plausible joint research project. Five such "speed collaboration" conversations would happen during a given session, ensuring new contacts were made for each participant.

- **"On the Job Market" Videos.** As mentioned, we created a *YouTube* channel associated with the *Halo21* program. As part of this, we invited junior researchers to produce a short 4-5 minute video describing themselves, their science with any papers, plots that they wanted to highlight. They could then submit their video to the organizers who would post it on the *YouTube* channel. Each week, we would highlight in our announcements these young scientists to the community and encourage people to watch them.

## 9) Encourage Diverse and Inclusive Participation

Workshops and conferences can benefit greatly from a diverse group of participants by utilizing their very different perspectives and contributions. Furthermore, without the physical limitations of an in-person meeting, virtual workshops are virtually unconstrained in terms of participation, both geographically and in terms of number of participants. Thus, it behooves the organizers of online conferences to be as inclusive as possible and provide opportunities for a large and diverse population of researchers to participate in the conference [4]. That said, care must be taken to invite members of under-represented groups to give talks and sit on panel discussions, so as to best showcase the breadth of perspectives coming from the community.

### Comments based on *Halo21*:

Our remote format encouraged the participation of astrophysical researchers that spanned the full range of career stages (undergraduate to faculty emeritus), from locations all over the world, and from a variety of backgrounds. All of the organizers advertised the workshop broadly among their collaborations, at their institutions, and via social media. In addition, we encouraged the many invited participants to invite others in the same manner, and generally encouraged the participation of everyone with an interest in gaseous halos.

Of the 17 invited keynote speakers, seven (41%) were women at the faculty level. Similarly, 40% of our invited expert panelists who helped to lead the post-keynote discussions were women. In addition, the organizers reached out to as many women as possible to fill the role of *Slack* conversation group leader, and in general succeeded in these efforts. While we were successful in achieving a reasonable gender balance among high-profile leadership and speaking roles at the workshop, we acknowledge that people of color were poorly represented in these roles. The vast majority of non-white participants in this workshop were either international and/or at the postdoctoral and graduate-student levels (or both; in particular, we had approximately 15 graduate students from Indian Institutes participating regularly). This underrepresentation of people of color affects all of the field of astronomy (and many other sciences), and is something all of the organizers are committed to addressing, not only in the context of this workshop.

## 10) Keep a Master Document with Content Updated Daily (i.e., Expect for People to Skip Parts of the Program)

In any workshop or conference, an easy-to-access schedule must be updated regularly for participants to know when and what to expect, but this updated schedule is crucial in the case of a virtual program where there are no in-person meetings to ensure everyone is coordinated.  Expect for your audience to not always be able to make every individual session, and enable a means for them to catch up on what they missed, by providing links to past content in a variety of forms (e.g., video recordings, *Slack* conversations, presentation slides).  This also allows for new members of the community to join mid-way through the program.  **We believe keeping this master document available and up-to-date is critical for the success of the overall program.**

### Comments based on *Halo21*:

We hosted a [*Halo21 Google* document](#) where we kept all relevant conference information.  This included a brief introduction to the workshop, format of the weekly programs, detailed schedule, *Zoom* links, *Slack* invite link, and links to the video recordings, transcripts, and presentation slides of each previous session.  This *Google* document was aliased to an easy-to-remember URL [http://bit.ly/KITP_CGM.](http://bit.ly/KITP_CGM)  Notably, this schedule was painstakingly updated every day with links to the content and recordings of that day, so that people who missed a session could easily catch up by watching the video of the *Zoom* recording, stepping through presentation slides, or reading over the transcript.  This helped to avoid the common pitfall of participants missing a few days and then never being able to rejoin the ongoing conversation.

## 11) Evaluate and Correct throughout the Program

Like in any experiment, better results come by actively incorporating feedback.  Organizers of virtual workshops should consider requesting feedback from participants and modifying aspects of the program accordingly over the course of the program itself.

### Comments based on *Halo21*:

At the end of each week of the program, we surveyed all participants to get feedback on the success of the various aspects of our workshop and seek improvements.  Since our virtual program lasted 8 weeks, it provided sufficient time to modify and tweak the format from week to week.  The organizers met weekly to decide on programmatic changes, which occurred over the first half of the program but then settled into an optimized format for the remainder.  In particular, one of the pieces of feedback we acted upon was the desire for more social networking activities, and for a more interactive program as opposed to one marked primarily by presentations.

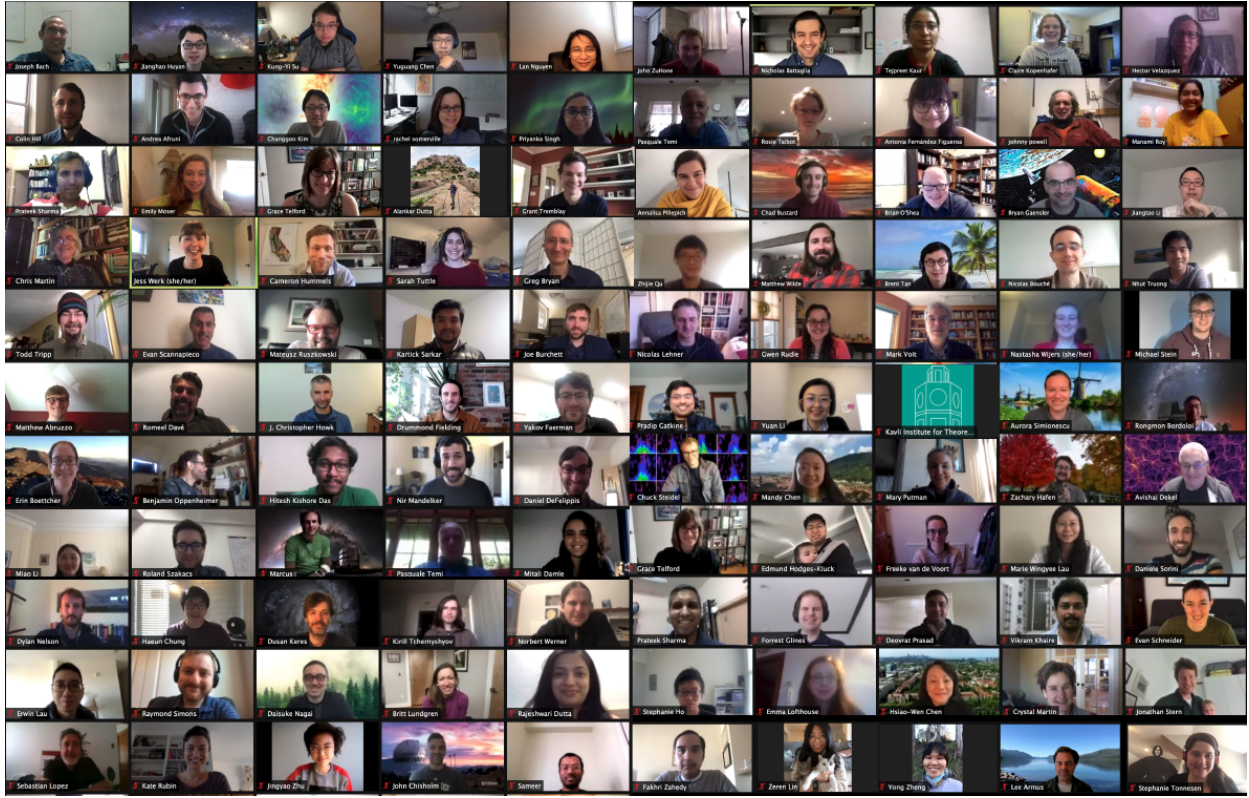

**Figure 1**
The conference photo from the *Halo21* workshop showing a subset of the total number of participants. Virtual programs can accommodate a very large number of attendees.

## Discussion and Feedback on the *Halo21* experiment

Historically, scientists generally view online meetings as inferior to in-person meetings. However, the feedback we received on *Halo21* was highly positive:

> *"I'd like to praise your organization of the KITP workshop. It was truly at a unique level of organization that I have not seen in other occasions. …. As a result, the participation was impressive and the science level extremely high."*

> *"It was the best conference that I have ever been to in nearly 50 years in terms of learning about an important and contemporary topic in physics."*

> *"This workshop has been the best organized and most useful conference I have ever been a part of, without question."*

> *"The organizers have done a tremendous job with this workshop. If there is some way you could share the overall format of your approach publicly, I think that could be very useful to the field as a whole as we move to hybrid online/in-person meetings."*

> *"One of the many important impacts of the workshop was starting a community-wide online discussion of the CGM. This forum would likely not have been possible if the meeting were held in-person—its creation is a rare opportunity."*
>
> *"I'm a first year PhD student, so this workshop was the perfect introduction to the field! I really appreciate how inclusive you made the conference. You did a great job committing to the virtual setting, using every available resource to make the workshop as effective as possible."*

Notably, both world-leading experts and first-year graduate students reported finding the program highly valuable. An extended KITP program typically provides world-leading experts with opportunities to learn from one another, forge new collaborations, and extend the frontiers of their field. *Halo21* managed to replicate those features of an in-person KITP program while also benefiting younger, less-experienced scientists, some of whom are just getting started.

Here are some thoughts on why *Halo21* worked so well:

- **Hundreds of participants.** The pandemic excused the organizers from one of the most painful tasks of KITP program organization—deciding who gets to spend a few weeks in "science heaven" and who does not, because of space limitations. Instead of limiting the program to just experts, we chose to include anybody from the CGM community who wanted to participate. The community response was astounding, particularly as word of mouth started to propagate during the program's first week. *Halo21* became "The Place To Be" for anybody working on CGM topics, without a travel or expense barrier. Consequently, the program benefited from participation by experts who would not have been able to attend in person as well as from participation by researchers whose recent findings were less well known to the organizers. The broad community participation in this program by hundreds of astrophysicists around the world was one of its primary successes. Figure 1 displays a conference photo from *Halo21* with only a fraction of the total number of participants.
- **All participants were remote and equal.** Because there was no in-person component of the *Halo21* program, all participants (including the organizers) were connecting virtually. This meant that everyone was on equal footing, and it did not set up a different tiers of participants. Organizers of hybrid or partially-remote meetings should consider the effect of having different classes of participants (i.e., some in-person and some remote), and how one may feel inferior as a remote participant if they aren't able to participate in an equal way. Care should be taken to ensure everyone is afforded the maximum opportunities for participation and engagement in the program.
- **Starving for interaction.** Our program also benefited from the pandemic in a way that future virtual programs might not be able to replicate—the community of participants was starving for scientific interaction with like-minded colleagues. They brought enormous enthusiasm, commitment, and energy to the program. Perhaps the CGM community would have brought the same energy to the program during more normal times. However, we got the impression that most of them were looking forward to 8-10 AM Pacific

Time each day of the workshop for that positive jolt of face time with colleagues that became all too rare during the depths of the pandemic.
- **Extended duration.** There were other virtual scientific meetings during the pandemic, but most were short, spanning just a few days, and some were packed so full of presentations that they brought on *Zoom* fatigue. In contrast, the eight-week duration of ***Halo21*** was a great blessing. It provided space for participants to digest what was developing and also allowed people with other commitments to disengage and then return to the discussions, according to their individual schedules. KITP's support with video recordings of talks and discussion sessions, along with *Slack* discussion threads, was greatly beneficial for providing continuity, so that intermittent participants could catch up on what was happening while they were away.
- **Audience members were only expected to participate for 2 hours each day.** Because each session lasted only 2 hours each day, those participating were expected to give their undivided attention during that short window. Using this short window each day helped prevent workshop burnout, while at the same time providing a reasonable amount of time to engage with the content. The program was greatly strengthened by the undivided attention that participants provided during this brief window each day.
- **Limited formal presentations.** Aside from the keynote talks, the only other formal presentations were the Wednesday tutorials, which numbered between two and four each week, for a total of four to six formal talks (lasting < 3 hours total) during a particular week of the program. That restriction made the program more discussion oriented than presentation oriented, encouraging more active participation than passive participation.
- **Contributed videos.** Recognizing that junior people need visibility in order for their careers to thrive, we created an alternative outlet for contributed presentations—the New Results videos on our ***Halo21*** *YouTube* channel. Those videos were generally of excellent quality and were very well received by the other participants. A strict duration limit of 5 minutes seems to be just right for getting the basic ideas of a new paper across and tempting others to seek the paper out for a deeper dive. Having those videos available in a *YouTube* channel enables asynchronous viewing, repeat viewing, and even the possibility of going viral.
- **Abundant and varied discussion opportunities.** Here is where the experimental spirit of the virtual workshop paid the greatest dividends. We tried everything we could think of to get people to talk to each other, across the usual disciplinary and generational boundaries. That may be the main reason so many people found the meeting unusually engaging and stimulating. Digital media assisted those ends by accommodating diverse communication styles and comfort levels. The greatest lasting impacts of our program are the collaborations it fostered and the lifting of entry barriers into our field for junior participants.
- **Both synchronous and asynchronous.** Incorporating a range of different synchronous and asynchronous participation methods was crucial to the success of the program. Confident, extroverted experts took advantage of the opportunities presented by panel discussions and the "raised hand" feature of *Zoom* sessions. More introverted and less confident participants could take their time and craft careful commentary on *Slack* threads. Digital recording and dissemination captured everything for people unable to participate in real time. The *Slack* conversations in particular became a diamond mine of recorded thoughts, observations, conversations, and contributions that program participants are still benefiting from.

Despite the successes of this virtual format, it does have some drawbacks:

- **A lot of work for organizers.** Good conferences take work: to plan, to organize, and to execute. The same is true for virtual workshops. It required a great deal of work to plan the layout of the meeting, invite speakers, and come up with relevant topics to engage the community. In particular, significant work went into recording and posting all workshop content each day to keep the ***Halo21*** master document up to date. This included posting presentation and new results videos to *YouTube*, including *Zoom* transcripts, gathering presentation slides from speakers and storing on *Google Drive*, and linking it all through the master document. All of this while trying to also participate in the relevant scientific discussions was certainly a handful, and we were glad to have 4 organizers to help balance the load.
- **Lacking in social interactions.** Despite the efforts we made towards increasing social interactions and networking through the program, it was still a challenge due to the distributed and isolated nature of virtual contact. For instance, it can be difficult to motivate someone to participate in a social event like pub trivia when it's 10AM local time. Now that the COVID-19 pandemic has passed, it may be difficult to motivate virtual social activities as it was in 2021.
- **Hard to have undivided attention of participants.** Scientists are busy with all manner of obligations between teaching, committee work, research, and real life. Motivating them to take time out of their day to focus on a fully virtual conference can be a hard ask. It will always be easier to get people to focus on an in-person workshop. The key is to anticipate that people may have to miss hours or days of a meeting, and provide ways that they can catch back up with what they have missed through recorded content (see #10).

Accounting for all of these considerations, we believe that virtual meetings can be very fruitful and productive. Even now as life has returned to normal after the COVID-19 pandemic, there is a unique role that virtual meetings can play in energizing a scientific community, potentially even moreso than traditional in-person meetings. We hope that other organizers will consider our advice in planning future meetings to incorporate an online / virtual component. We encourage interested readers to further delve into the details of ***Halo21*** by perusing our master scheduling document: http://bit.ly/KITP_CGM.

Acknowledgements
*This research was supported in part by grant NSF PHY-2309135 to the Kavli Institute for Theoretical Physics (KITP).*

# References

- Hodapp, Theodore, and Erika Brown. "Making Physics More Inclusive." Nature, vol. 557, no. 7707, May 2018, pp. 629–32. www.nature.com, https://doi.org/10.1038/d41586-018-05260-4.

    ↩

- Keengwe, Jared, and Gary Schnellert. "Fostering Interaction to Enhance Learning in Online Learning Environments." IJICTE vol.8, no.3 2012: pp.28-35. https://doi.org/10.4018/jicte.2012070104
↩
- Qi, Fan, et al. "Facilitating Interdisciplinarity: The Contributions of Boundary-Crossing Activities among Disciplines." Scientometrics, vol. 129, no. 10, Oct. 2024, pp. 6435–53. DOI.org (Crossref), https://doi.org/10.1007/s11192-023-04924-x.
↩
- Skiles, Matthew, et al. "Conference Demographics and Footprint Changed by Virtual Platforms." Nature Sustainability, vol. 5, no. 2, Feb. 2022, pp. 149–56. www.nature.com, https://doi.org/10.1038/s41893-021-00823-2.
↩